\documentclass[
aps,%
12pt,%
final,%
notitlepage,%
oneside,%
onecolumn,%
nobibnotes,%
nofootinbib,
superscriptaddress,%
noshowpacs,%
centertags,
prd,amssymb,showpacs,floatfix,preprint,nofootinbib,superscriptaddress]%
{revtex4}

\usepackage{upgreek,amsmath,amsfonts,graphics,epsfig,amssymb,color, tocbibind}

\renewcommand\bibname{\normalfont\centerline{REFERENCES}\global\def\bibname{REFERENCES}}

\usepackage[figurename=Fig.]{caption}
\usepackage[labelsep=period]{caption}

\begin{document}

\begin{center}
\large \bfseries {Potassium influence on Earth's mantle convection and Borexino data}

 {\bf \copyright~2025 I.\,S.\,Karpikov$^{\bf 1,}$}{\renewcommand{\thefootnote}{*}\footnote{E-mail: karpikov@inr.ru}},
    \\
\end{center}

\begin{center}    
{  $^1$Institute for Nuclear
Research of the Russian Academy of Sciences, 60th October Anniversary
Prospect 7a, Moscow 117312, Russia }\\

\end{center}
 
\

\begin{center}
\begin{minipage}{\textwidth }
\small
{\bf Abstract}


High flux of geoantineutrinos $^{40}$K and geoneutrinos $^{40}$K ($^{40}$K-geo-($\bar{\nu} + \nu$)) can be obtained from reanalysis of the Borexino Phase III data. Large amounts of $^{40}$K should produce a significant heat flow that should affect Earth's internal processes. We present the results of the modeling of mantle convection taking into account the excess heat from $^{40}$K.  

\end{minipage}
\end{center}

%
\newpage
\begin{center}
1. INTRODUCTION
\end{center}

Research of the internal structure of the Earth is extremely important for understanding the nature of seismic activity. Of great interest is also the question of the influence of the Earth's internal heat on climate. The Borexino underground experiment \cite{bib:Nature2018} is primarily designed to study solar neutrinos, but one of its scientific tasks is also the search for geoneutrinos, which are formed during the decay of uranium and thorium isotopes \cite{bib:Borexino_geo}. The Borexino paper \cite{bib:Borexino_CNO} presents an improved analysis of the measurement of solar neutrinos from the Carbon-Nitrogen-Oxygen (CNO) cycle. From the results of the analysis of the Phase III data set, the measured rate $R_{\rm CNO} = 6.7^{+2.0}_{-0.8}$ cpd / 100t (counts / (day $\times$ 100 tonnes)). 

However, the spectrum of solar neutrinos from the CNO cycle is very similar to the spectrum of geo-antineutrino and neutrino from $^{40}$K decay ($^{40}$K-geo-($\bar{\nu} + \nu$)). The works \cite{bib:Bezrukov_Sinev, bib:Bezrukov_Sinev_Karpikov} demonstrated an analysis of the recoil electron spectrum taking into account the neutrino flux $^{40}$K-geo-($\bar{\nu} + \nu$). Indeed, adding the spectrum of $^{40}$K-geo-($\bar{\nu} + \nu$) recoil electrons gives a better fit than if the data only included the spectrum of the neutrino flux from the CNO cycle. The work \cite{bib:Bezrukov_Low_metall} also shows that the improved fit of the spectrum with $^{40}$K-geo-($\bar{\nu} + \nu$) gives a less expected amount of neutrinos from the CNO cycle, which in turn indicates a low metallicity of the Sun. These articles demonstrated the rate  R($^{40}$K) = 11 cpd/100t. Thus, the mass fraction of potassium is 3.2\% \cite{bib:News_RAS, bib:Earth_electric_field} of the total mass of the Earth. The average potassium content in the Earth's crust is 2. 4\% \cite{bib:40K_in_crust}.  

The average energy release in decay of $^{40}$ K is $E_{release}$ = 0.6 MeV. With such a concentration of potassium in the Earth's mantle, a lot of radiogenic heat should be released. In turn, this heat will significantly affect the processes inside the Earth. The purpose of this work is to evaluate the real possibility of the existence of a large potassium content in the Earth, taking into account the geophysical model of mantle convection. 

\begin{center}
2. TWO-DIMENSIONAL CONVECTION MODELS 
\end{center}
\begin{center}
2.1 Governing Equations
\end{center}

To describe the flows in the upper mantle of the Earth, we solved a system of dimensionless governing equations of mass, momentum, and energy
conservation in a two-dimensional spherical ring geometry in the Boussinesq approximation \cite{bib:Ricard}.
\begin{equation}
  -\nabla P + \nabla \cdot [\eta(\nabla\mathbf{v}+(\nabla\mathbf{v})^{T})]=RaT\mathbf{e_{r}}
\label{Eq:Eler}  
\end{equation}  
\begin{equation}
\nabla \cdot \mathbf{v}=0
\label{Eq:divV}
\end{equation}  
\begin{equation}
\frac{\partial T}{\partial t}+(\mathbf{v}\cdot\nabla)T=\nabla\cdot(\kappa\nabla T)+ H(t)
\label{Eq:term}
\end{equation}  
where $\mathbf{v}$ is the velocity vector; $P$ is the dynamic pressure; $T$ is the temperature; $t$ is the time; $\eta$ is the coefficient of dynamic viscosity; $\kappa$ is the coefficient of thermal diffusivity; $H(t)$ is an internal heating source. $Ra$ is the Rayleigh number, defined as
$Ra = \alpha  \rho_{0}  g  \Delta T D^{3}/(\eta_{0} \kappa_{0})$ 
with $\alpha$ the surface thermal expansivity; $g$ is gravitational acceleration; $\Delta T$ is the temperature between the mantle-core boundary and the mantle-crust boundary; $\rho$ is the reference density; $\eta_{0}$ is the reference viscosity; $D$ is the thickness of the mantle.

For non-dimensionalization, we use scaling factors: $x=X/D$, $y=Y/D$, $t=t^{'}\kappa_{0}/D^{2}$, $T=T^{'}/\Delta T$, $\mathbf{v}=\mathbf{V}D/\kappa_{0}$, $P=P^{'}D^{2}/(\kappa_{0} \eta_{0})$.  In this paper, for qualitative evaluation, we neglect the dependence of $\eta$ and $\kappa$ on temperature and statistical pressure and take them equal to one.

We accept the initial conditions for temperature $T_{0}=\Delta T$ and for velocity $\mathbf{v_{0}}=0$. The boundary conditions for the velocity are no leakage, that is, the velocity at the limits is zero $\mathbf{v|_{r=0}}=\mathbf{v|_{r=D}}=0$.  The boundary conditions for temperature: mantle-crust boundary $T|_{r=0}=0.12 \Delta T$ and mantle-core boundary $T|_{r=D}=1.12 \Delta T$. The dynamic pressure at the mantle-crust boundary is zero.

\begin{center}
2.2 Simulation parameters
\end{center}

The parameter values for the mantle were taken from \cite{bib:Schubert, bib:Tosi} and are the following: 
 $\alpha_{0}=2\times10^{-5}$~K$^{-1}$,
 $\rho_{0}=4.5\times10^{3}$~kg/m$^{3}$,
 $D=2900$ km,
 $\Delta T=3500$K,
 $\eta_{0}=5\times10^{21}$ Pa$\cdot$s,
 $\kappa_{0}=10^{-5}~$m$^{2}$/s.

The paper \cite{bib:Arnould, bib:Trubitsyn} presents various models of mantle convection with Rayleigh numbers $Ra$ in the range of 10$^{5}$ to 5$\times$10$^{7}$ and internal heating sources $H$ in the range from 8.7$\times$10$^{-13}$ to 1.4$\times$10$^{-11}$ W/kg.  In our work, we will consider four models with the Rayleigh number $Ra$ = 10$^{5}$ and $Ra$ = 10$^{6}$; Internal heating source for today $H_{now}$ = 5$\times$10$^{-12}$ W/kg and $H_{now}$ = 15$\times$10$^{-12}$ W / kg (if the mantle density is $\rho_{0}=4.5\times10^{3}$~ kg / m$^{3}$ then the total heat flow is 20 TW and 60 TW, respectively). Large Rayleigh numbers and large internal heat lead to unstable turbulent solutions, and we do not consider them for now. 

Since the amount of potassium decreases exponentially with time, the internal heating source in the past was greater:
\begin{equation}
H_{0} = H_{now}\times \exp(t_{max}/t_{40K})
\end{equation}
where $t_{max}$ = 4.5 billion years is the lifetime of Earth, $t_{40K}$ = 1.8 billion years is the lifetime of $^{40}$K. As a result, internal heat depends on the time as follows. 
\begin{equation}
H(t) = H_{0}\times \exp(-t/t_{40K})
\end{equation}

We normalize internal heating sources to the value $H_{norm} = (c_{p}\kappa_{0} \Delta T)/D^{2}$, where $c_{p} = 1.25 \times10^3$ J/(kg K) is the heat capacity, $H_{norm}$ = 6.3$\times$10$^{-13}$ W/kg.

\begin{center}
3. RESULTS OF NUMERICAL MODELING OF MANTLE CONVECTION
\end{center}

To solve the system of governing equations, we used the Wolfram Mathematica 14 package. We used 1000 time steps, which is 4.5 Myr. The size of the spatial cell is 14.5 km. If necessary, the author can share his calculations.

\begin{figure}
\begin{minipage}[h]{0.47\linewidth}
\center{\includegraphics[width=1\linewidth]{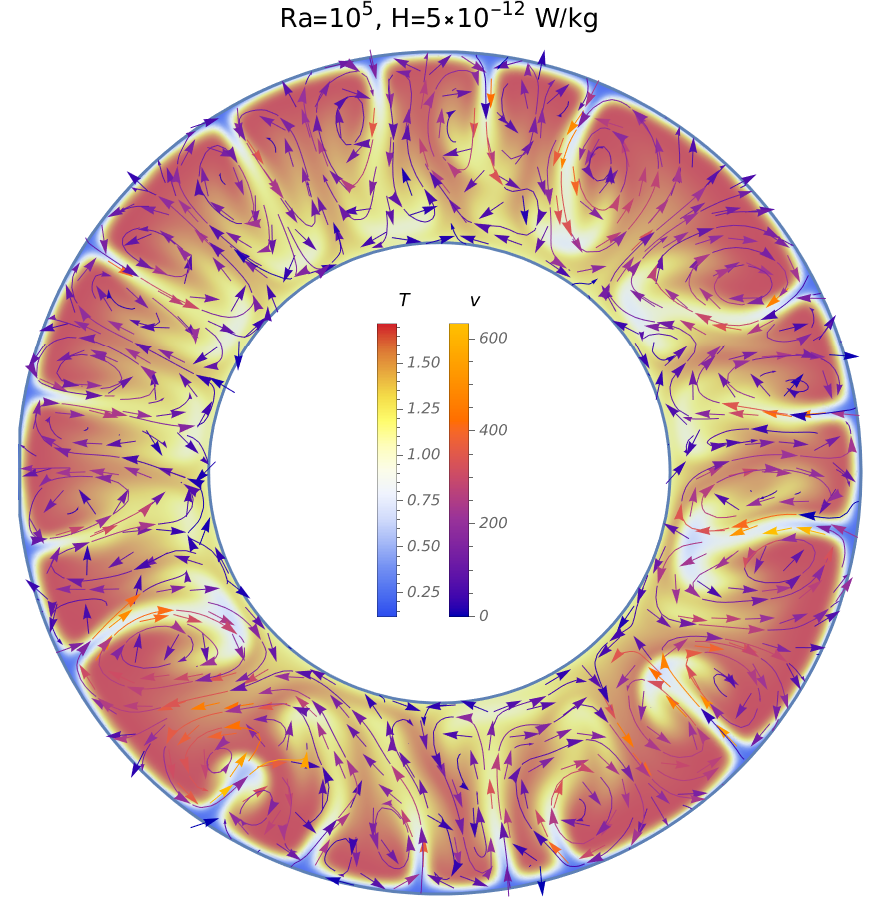}} a) \\
\end{minipage}
\hfill
\begin{minipage}[h]{0.47\linewidth}
\center{\includegraphics[width=1\linewidth]{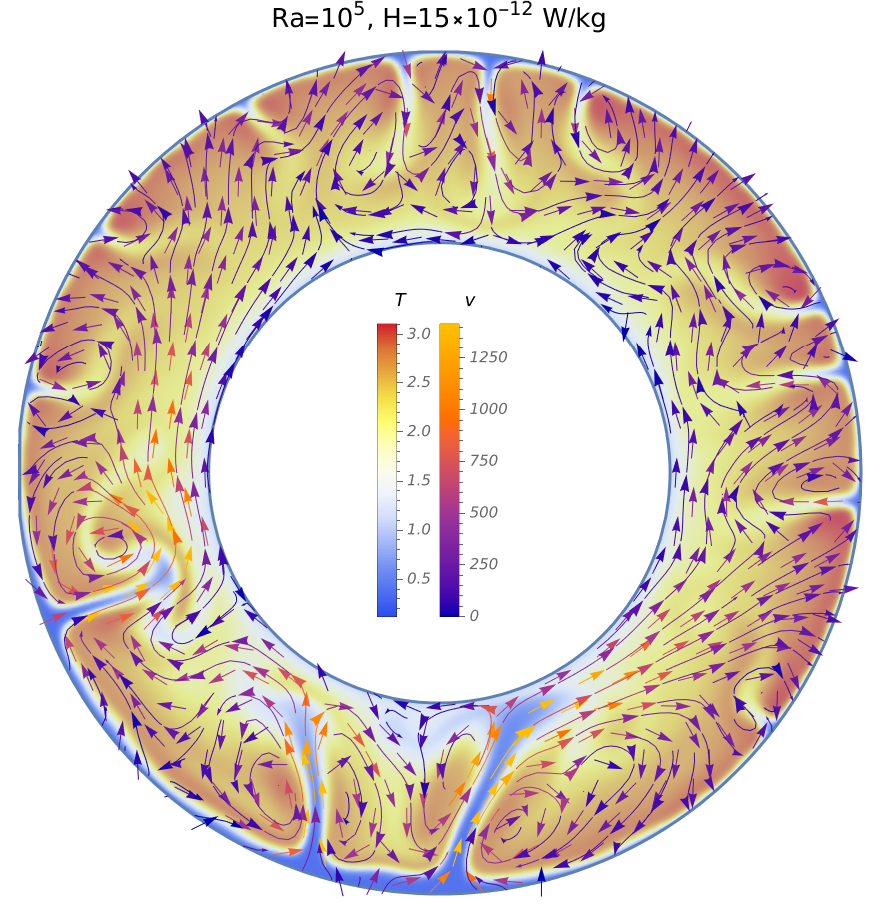}} \\b)
\end{minipage}
\vfill
\begin{minipage}[h]{0.47\linewidth}
\center{\includegraphics[width=1\linewidth]{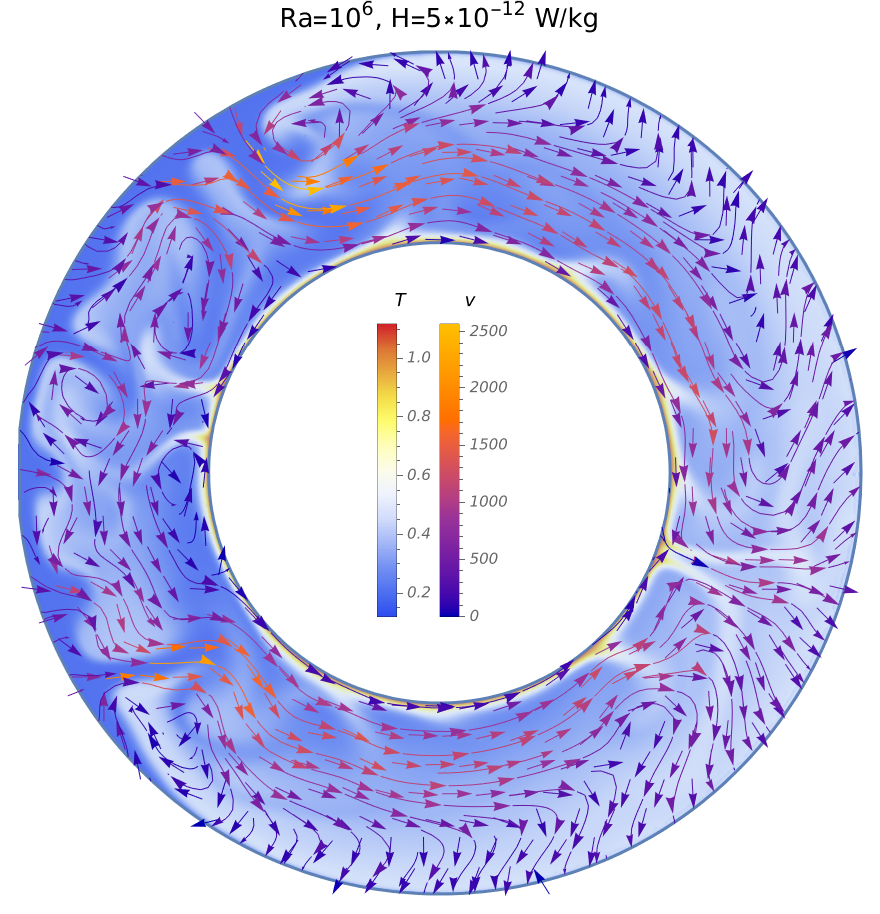}} c) \\
\end{minipage}
\hfill
\begin{minipage}[h]{0.47\linewidth}
\center{\includegraphics[width=1\linewidth]{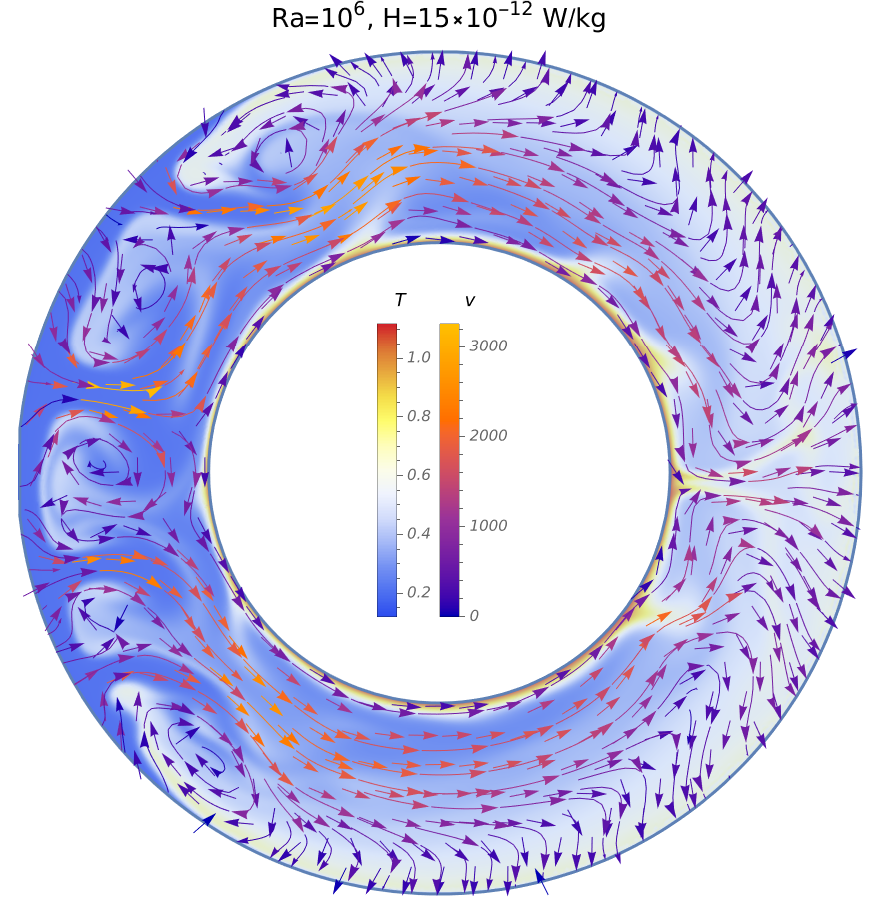}} d) \\
\end{minipage}
\caption{The result of solving the governing equations \ref{Eq:Eler}, \ref{Eq:divV}, \ref{Eq:term}. Two-dimensional distribution of dimensionless temperatures and velocities in the mantle. Temperatures are shown by color, velocities indicated by arrows. 
a) Rayleigh number $Ra=10^{5}$, internal heating source for today $H_{now}=5 \times10^{-12}$ W/kg; b) Rayleigh number $Ra=10^{5}$, internal heating source for today $H_{now}=15 \times10^{-12}$ W/kg; c) Rayleigh numbe $Ra=10^{6}$, internal heating source for today $H_{now}=5 \times10^{-12}$ W/kg; d) Rayleigh number $Ra=10^{6}$, internal heating source for today $H_{now}=15\times10^{-12}$ W/kg }
\label{fig:2D}
\end{figure}

Figure \ref{fig:2D} shows the temperature and velocity distributions for different mantle models. From this figure it is clearly seen that the model with high viscosity ($Ra=10^{5}$) heats up very strongly and the Earth does not have time to cool down today. Even a model with a small amount of heat from $^{40}$K ($H_{now}=5 \times 10^{-12}$ W/kg) shows a very strong warming of the Earth. However, the model with lower viscosity ($Ra=10^{6}$) cools faster because the thermal energy is converted into the kinetic energy of the mantle flow. 

\begin{figure}
\begin{minipage}[H]{0.7\linewidth}
\center{\includegraphics[width=1.0\linewidth]{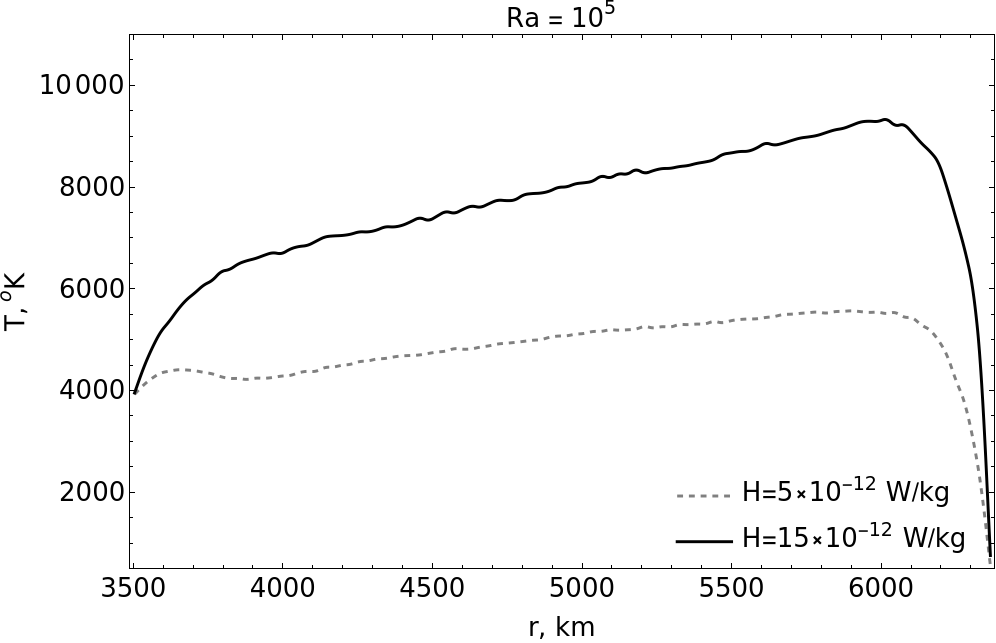}}
\end{minipage}
 \vfill
\begin{minipage}[H]{0.7\linewidth}
\center{\includegraphics[width=1.0\linewidth]{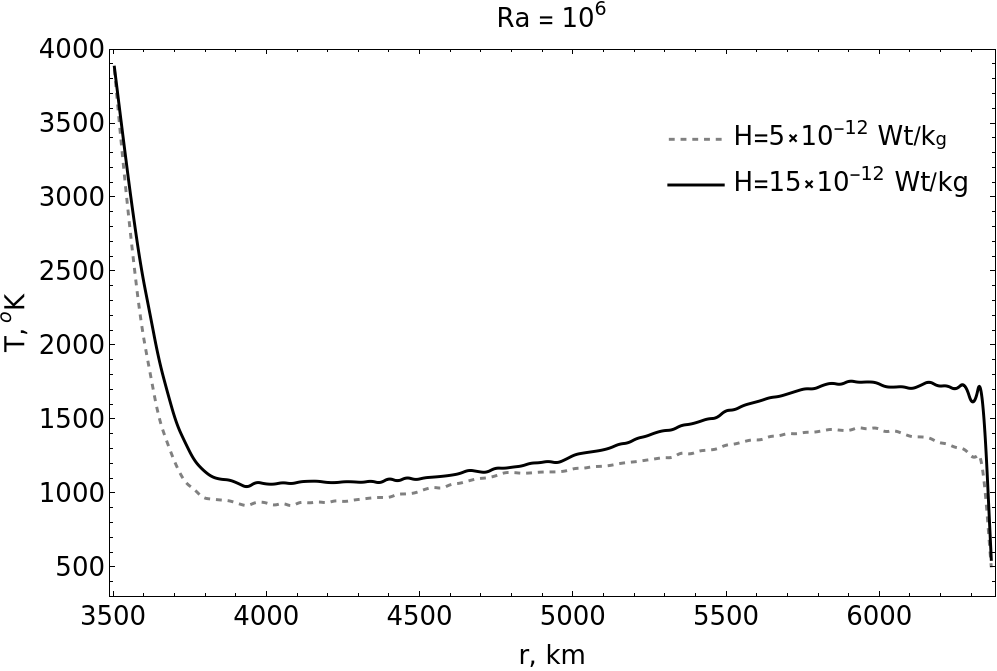}}
\end{minipage}
\caption{Dependence of the average temperature on the distance from the center of the Earth. Top: Model with high viscosity, Rayleigh number $Ra=10^{5}$.
Bottom: Model with lower viscosity, Rayleigh number $Ra=10^{6}$.}
\label{fig:T_vs_r}
\end{figure}

\begin{figure}
\begin{minipage}[h]{0.7\linewidth}
\center{\includegraphics[width=1.0\linewidth]{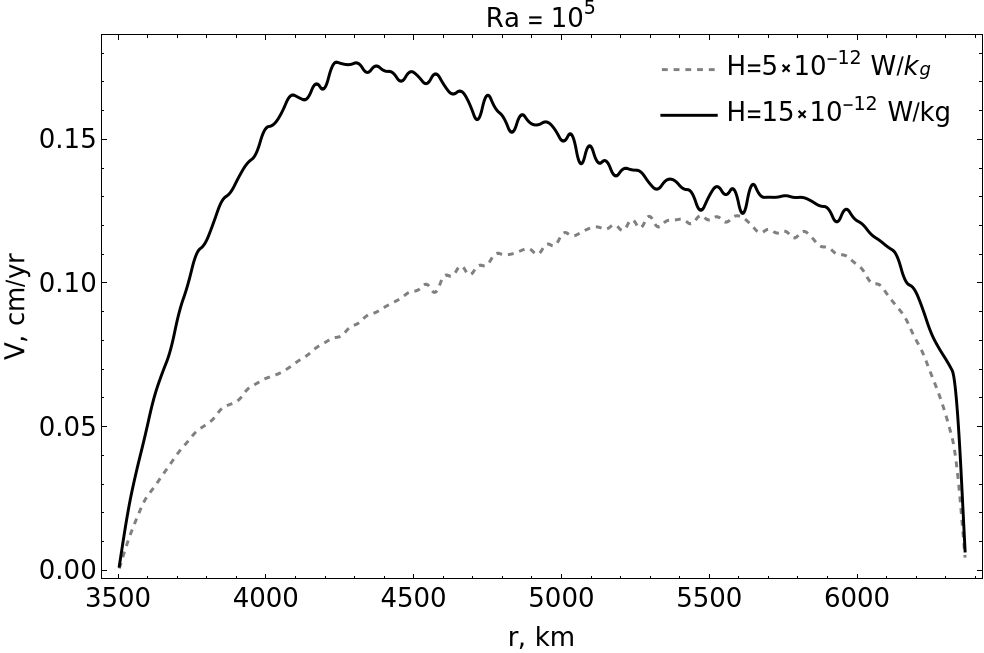}}
\end{minipage}
\vfill
\begin{minipage}[h]{0.7\linewidth}
\center{\includegraphics[width=1.0\linewidth]{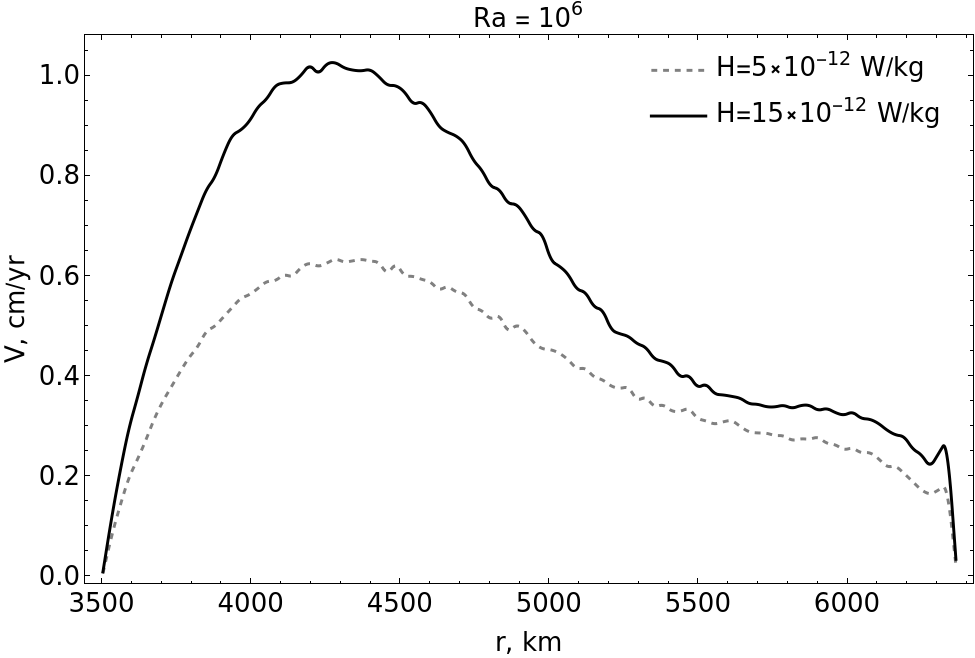}}
\label{fig:Flux_E}
\end{minipage}
\caption{Dependence of the average velocity module on the distance from the center of the Earth. Top: Model with high viscosity, Rayleigh number $Ra=10^{5}$.
Bottom: Model with lower viscosity, Rayleigh number $Ra=10^{6}$.}
\label{fig:V_vs_r}
\end{figure}

Figure \ref{fig:T_vs_r} shows the dependence of the average temperature on the distance from the center of the Earth. This figure clearly shows that in the high viscosity model, the heat from $^{40}$K heated the mantle to temperatures above the core temperature, but in the low viscosity model the mantle had time to cool down. Figure \ref{fig:V_vs_r} shows the dependence of the average velocity module on the distance from the center of the Earth.

\begin{center}
4. CONCLUSIONS AND OUTLOOK
\end{center}

From the results of the Borexino data analysis presented in the works \cite{bib:Bezrukov_Sinev, bib:Bezrukov_Sinev_Karpikov, bib:Bezrukov_Low_metall}, it follows that there is a large $^{40}$K content on the Earth. In our work, we demonstrated how realistic this result is and how consistent it is with the models of the Earth's structure. In our work, we have qualitatively demonstrated that, under certain mantle parameters, such as viscosity, a high potassium content in the Earth's mantle is possible. Indeed, we have shown that at high Rayleigh numbers the cooling of the Earth occurs much faster. The results of the Borexino data analysis can help to better understand and clarify the structure of the Earth, so it is very important that similar experiments be carried out in the future. 

 In the future, we plan to further develop the numerical model of convection of the earth's mantle taking into account the heat from potassium 40. Our plans include in the modeling the dependence of viscosity and thermal diffusivity on temperature and static pressure.

\begin{center}
Acknowledgments
\end{center}

We are grateful Anton Bernatskii, Denis Kuleshov, Dmitry Kirpichnikov, Petr Satunin, and Valery Sinev for an interesting discussion. We also express our gratitude to Petr Tenishev and the Russian community Wolfram Mathematica. 

\newpage

\begin{center}
FIGURE CAPTIONS
\end{center}

{\bf Fig.1.} Two-dimensional distribution of dimensionless temperatures and velocities in the mantle. Temperatures are shown by color; velocities are indicated by arrows. 
a) Rayleigh number $Ra=10^{5}$, internal heating source for today $H_{now}=5 \times10^{-12}$ W/kg; b) Rayleigh number $Ra=10^{5}$, internal heating source for today $H_{now}=15 \times10^{-12}$ W/kg; c) Rayleigh numbe $Ra=10^{6}$, internal heating source for today $H_{now}=5 \times10^{-12}$ W/kg; d) Rayleigh number $Ra=10^{6}$, internal heating source for today $H_{now}=15\times10^{-12}$ W/kg 

{\bf Fig.2.} Dependence of the average temperature on the distance from the center of the Earth. Top: Model with high viscosity, Rayleigh number $Ra=10^{5}$.
Bottom: Model with lower viscosity, Rayleigh number $Ra=10^{6}$.

{\bf Fig.3.} Dependence of the average velocity module on the distance from the center of the Earth. Top: Model with high viscosity, Rayleigh number $Ra=10^{5}$.
Bottom: Model with lower viscosity, Rayleigh number $Ra=10^{6}$.

\newpage

\newpage

\clearpage

\newpage

\clearpage

\newpage

\newpage

\end{document}